 \documentclass[preprint,12pt]{elsarticle}

\usepackage{graphicx}

\usepackage{amssymb,amsbsy,amsmath}

\usepackage{color}
\usepackage{amsbsy,amsmath,amssymb}
\usepackage{upgreek}
\usepackage{braket}
\usepackage{textcomp}

\journal{Optics Communication}

\begin{document}
\begin{frontmatter}



\title{Non-integer OAM beam shifts of Hermite-Laguerre-Gaussian beams}

\author{A.M. Nugrowati\corref{cor1}}
\ead{nugrowati@physics.leidenuniv.nl}
\cortext[cor1]{Corresponding author}
\author{J.P. Woerdman}
\address{Huygens Laboratory, Leiden University\\ P.O. Box 9504, 2300 RA Leiden, The Netherlands}

%


\begin{abstract}
We have studied the effect of non-integer Orbital Angular Momentum (OAM) on OAM enhanced beam shifts, for in-plane (Goos-H\"{a}nchen) and out-of-plane (Imbert-Fedorov) shifts, using Hermite-Laguerre-Gaussian
beams. Contrary to naive expectation we find, theoretically and experimentally, that the non-integer OAM beam shifts do \textit{not} interpolate linearly between the integer OAM beam shifts. 
\end{abstract}


\end{frontmatter}

\section{Introduction}
\label{sec:intro}
Beams having a finite transverse extent are known to be deflected differently from the geometric prediction, upon interaction with interfaces \cite{Goos:AnnPhys1947, Fedorov:1955, Imbert:PRD1972, Bliokh:Jopt2013}. Recently, an extensive research has been carried out addressing how this interaction is affected by the Orbital Angular Momentum (OAM) of the beam, which so far has been focussed on the integer OAM case. Specifically, OAM couples to the Goos-H\"{a}nchen (GH) shift \cite{Goos:AnnPhys1947} that is a longitudinal shift and in-plane with the incoming light, and the Imbert-Fedorov (IF) shift \cite{Fedorov:1955, Imbert:PRD1972} that is transversal and out-of-plane of incidence. Each of these two shifts can be separated into (i) the spatial type of shift ($\Delta_\mathrm{GH}$, $\Delta_\mathrm{IF}$) that is independent of beam focussing, and (ii) the angular type of shift ($\Theta_\mathrm{GH}$, $\Theta_\mathrm{IF}$) that is enhanced by beam propagation upon focussing \cite{Aiello:OptLett2008, Bliokh:OpLett09}. The angular shift is closely related to the Spin Hall Effect of Light \cite{Hosten:Science2008, Qin:OptLetter2009, Hermosa:OpLett2011}. In this Communication our interest is in partial (e.g. external) reflection of OAM beams. In this case, it has been reported \cite{Gupta:OptCommun2006, Aiello:OptLett2008, Bliokh:OpLett09, Merano:PRA10, Hermosa:SPIE2011, Sasada:JOpt2013} that the shifts are linearly enhanced by the (integer) OAM content of the beam. For completeness we note that in the case of total (internal) reflection the beam shifts are purely spatial and are not affected by the OAM value of the beam \cite{Merano:PRA10, Loeffler:JOpt2013}.

\begin{figure}[htbp]
\begin{center}
\includegraphics{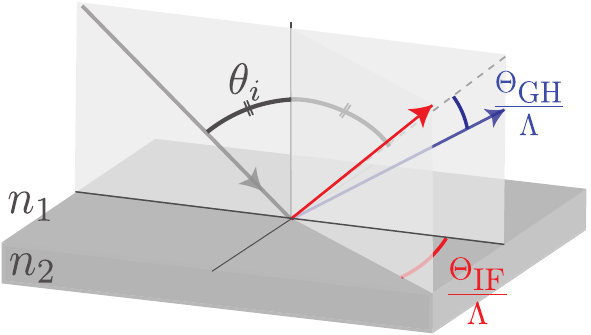}
\end{center}
\caption{\label{Fig01}Schematic of angular Goos-H\"{a}nchen ($\Theta_\mathrm{GH}$) and Imbert-Fedorov ($\Theta_\mathrm{IF}$) shifts normalized to the wavenumber $k_0$ and Rayleigh length $z_\mathrm{R}$ ($\Lambda=k_0 z_\mathrm{R}$).  We deal with partial external reflection, i.e. refractive index $n_1$\textless$n_2$.}
\end{figure}

In OAM beam generation one uses typically Laguerre-Gaussian (LG) beams  since their (integer) azimuthal mode index $\ell$ is directly proportional to the OAM value \cite{Woerdman:OptCommun1993}. Such integer OAM beams have a cylindrically symmetric beam profile. Our interest is in beam shifts of \textit{non-integer} OAM beams; such beams have anisotropic beam profiles. Several techniques have been introduced to generate non-integer OAM beams \cite{Oemrawsingh:PRL2005, Dwyer:OpEx2010, Courtial:OptCommun2004, Abramochkin:JOA2004}, mainly motivated by their potential in carrying higher density classical and quantum information. A light beam carrying a non-integer OAM value can be made in such a way \cite{Nugrowati:GenDetFOAM2012, Merano:PRA10, Abramochkin:JOA2004} that the (anisotropic) beam profile is preserved upon propagation, apart from the overall scaling and a quadratic phase factor. This is crucial for their use in beam shift experiments since it eliminates the dependence on the position of the reflecting (or transmitting) interface. In this Communication, we report our study of beam shifts of a structurally propagation invariant beam carrying non-integer OAM, upon external reflection on a dielectric surface. We find, theoretically and experimentally, that the angular type of beam shifts (shown in Fig.~\ref{Fig01}) has a strong nonlinear dependence on the non-integer OAM value, that is different from the linear dependence in the case of integer OAM \cite{Gupta:OptCommun2006, Hermosa:SPIE2011, Sasada:JOpt2013}.

 \section{Theory}
 \label{sec:theory}
We start with an input Hermite-Gaussian (HG$_{n,m}$) beam having the mode indices $n,m$ entering a `$\pi/2$-mode converter' \cite{Nugrowati:GenDetFOAM2012}. By varying the orientation angle $\alpha$ of its transverse profile with respect to those of the converter (see Fig.~\ref{Fig02} below), the output beam is a propagation invariant Hermite-Laguerre-Gaussian (HLG$_{n,m}\left(x,y|\alpha\right)$) beam carrying an arbitrary non-integer OAM value; i.e. $\ell=(m-n)\sin2\alpha$ \cite{Abramochkin:JOA2004}. At $\alpha=\mathrm{N}\pi/2$ and $\alpha=(2\mathrm{N}+1)\pi/4$ where $\mathrm{N}$ is an integer, the output beams are HG$_{n,m}$ and LG$_{p,\ell}$ modes, respectively, with $\ell=(m-n)$ and $p=\mathrm{min}(m,n)\mathrm{,}~p\geqslant0$ the radial mode index.

To derive the beam shifts for non-integer OAM beams, we follow the general expression of a HLG mode given by Abramochkin and Volostnikov \cite{Abramochkin:JOA2004},
\begin{multline}
\label{eq:HLG}
\mathrm{HLG}_{n,m}(x,y|\alpha)=e^{-x^2-y^2}\sum^{n+m}_{k=0}i^k\cos^{n-k}\alpha\sin^{m-k}\alpha\\P_k^{(n-k,m-k)}(-\cos2\alpha)H_{n+m-k}(\sqrt{2}x)H_k(\sqrt{2}y),
\end{multline}
with $P$ the Jacobi and $H$ the Hermite polynomials. We further use the expression given by Aiello \cite{Aiello:NJP2012} for the observable GH and IF shifts ($\braket{x}$ and $\braket{y}$, respectively) of general beams. The contribution of the angular and the spatial shifts of the fundamental TEM$_{00}$ mode, denoted by the superscript "0", to the total shifts is given by 
\begin{equation}
\label{eq:ObsShift}
\left[\begin{matrix}
\braket{x}  \\ \braket{y} 
\end{matrix}
\right ] =
\left[
\begin{matrix}
\Delta^0_\mathrm{GH}\\\Delta^0_{\mathrm{IF}}
\end{matrix}
\right]
+
\left[
\begin{matrix}
a_{11} & a_{12} \\ a_{21} & a_{22}
\end{matrix}
\right]
\left[
\begin{matrix}
\Theta^0_\mathrm{GH}\\\Theta^0_{\mathrm{IF}}
\end{matrix}
\right],
\end{equation}
where the dimensionless shifts of the TEM$_{00}$ mode are
\begin{subequations}
\begin{align}
\Delta^0_\mathrm{GH}&=w_\perp \mathrm{Im} \left(\dfrac{\partial\ln r_\perp}{\partial\theta}\right)+w_\parallel \mathrm{Im} \left(\dfrac{\partial\ln r_\parallel}{\partial\theta}\right),\\
-\Theta^0_\mathrm{GH}&=w_\perp \mathrm{Re} \left(\dfrac{\partial\ln r_\perp}{\partial\theta}\right)+w_\parallel \mathrm{Re} \left(\dfrac{\partial\ln r_\parallel}{\partial\theta}\right),\\
\begin{split}
\Delta^0_\mathrm{IF}&=-\dfrac{a_\parallel a_\perp \cot\theta}{R^2_\parallel a^2_\parallel+R^2_\perp a^2_\perp}\\ &\left[\left(R^2_\parallel+R^2_\perp\right)\sin \eta+2R_\parallel R_\perp \sin\left(\eta-\varphi_\parallel+\varphi_\perp\right)\right],
\end{split}\\
\Theta^0_\mathrm{IF}&=-\dfrac{a_\parallel a_\perp \cot\theta}{R^2_\parallel a^2_\parallel+R^2_\perp a^2_\perp}
\left[\left(R^2_\parallel+R^2_\perp\right)\cos \eta \right],
\end{align}
\end{subequations}
with $w_{\parallel/\perp}=R^2_{\parallel/\perp}a^2_{\parallel/\perp}/(R^2_\parallel a^2_\parallel+R^2_\perp a^2_\perp)$, $r_{\parallel/\perp}=R_{\parallel/\perp}\exp\left(i\varphi_{\parallel/\perp})\right)$ is the Fresnel reflection coefficient at incident angle $\theta$, $a_{\parallel/\perp}$ the electric field components, and $\eta$ their phase difference. The matrix elements in Eq.(\ref{eq:ObsShift}) are expressed as 
\begin{equation}
\label{eq:matrix}
a_{ij}=\dfrac{2~\mathrm{Im}\displaystyle\int{f^*\left(\beta_{i}\dfrac{\partial}{\partial \beta_{j}}\right)f \mathrm{d}x\mathrm{d}y}}{\displaystyle\int{|f|^2\mathrm{d}x\mathrm{d}y}},
\end{equation} 
with $f$ the general beam expression and $i,j$ the indices of the matrix elements ($\beta_{1}=x$, $\beta_{2}=y$). This matrix has been derived by Aiello for \textit{integer} OAM beam shifts \cite{Aiello:NJP2012}. 

For completeness, we would like to point out that the matrix $a_{i,j}$  is the same as the 4-by-4 matrix for the dimensionless OAM beam shifts in Merano et al. \cite{Merano:PRA10} with the GH shift $k_0\braket{x}=\Delta^\ell_\mathrm{GH}+\left(z/z_\mathrm{R}\right)\Theta^\ell_\mathrm{GH}$ and the IF shift $k_0\braket{y}=\Delta^\ell_\mathrm{IF}+\left(z/z_\mathrm{R}\right)\Theta^\ell_\mathrm{IF}$, having the OAM content $\ell$, the wavenumber $k_0$, and the Rayleigh length $z_\mathrm{R}$. Note the $z$ dependency of the angular shifts \cite{Aiello:OptLett2008}.

We introduce now a simple method to derive \textit{non-integer} OAM beam shifts: we decompose the HLG mode into two integer OAM beams with opposite signs (i.e. two LG modes with $\pm\ell$ signs). For clarity we restrict ourselves to the beam shifts of the HLG$_{1,0}(x,y|\alpha)$ mode but our approach is generally true for arbitrary HLG$_{n,m}$. By varying $\alpha$, we realize any non-integer OAM value between $-1\leqslant\ell\leqslant1$. The decomposition of the mode is
\begin{multline}
\label{eq:HLG10}
\mathrm{HLG}_{1,0}(x,y|\alpha)=\dfrac{\cos\alpha+\sin\alpha}{\sqrt{2}} \mathrm{LG}_{0,1}(x,y) \\+ \dfrac{\cos\alpha-\sin\alpha}{\sqrt{2}} \mathrm{LG}_{0,-1}(x,y),
\end{multline}
where LG$_{0,\pm1}(x,y)=e^{-x^2-y^2}(x\pm iy)^{\pm1}L_0^{\pm1}(2x^2+2y^2)$ and $L$ the Laguerre polynomial. 
Finally, substituting Eq.(\ref{eq:HLG10}) in Eq.(\ref{eq:matrix}) yields the matrix elements 
\begin{subequations}
\label{eq:matrixniOAM}
\begin{align}
\label{m11}
a_{11}&=\left(1+|m-n|\right)\left(2+\cos2\alpha\right)/2, \\ 
\label{m12}
a_{12}&=-|m-n|\sin2\alpha,\\
\label{m21}
a_{21}&=|m-n|\sin2\alpha, \\ 
\label{m22}
a_{22}&=\left(1+|m-n|\right)\left(2-\cos2\alpha\right)/2,
\end{align}
\end{subequations}
of our non-integer OAM enhanced beam shifts.

The antidiagonal matrix elements ($a_{12}$ and $a_{21}$) corresponding to the spatial beam shifts, behave similarly to the integer case \cite{Merano:PRA10, Hermosa:SPIE2011} in the sense that they are \textit{linearly} dependent on the non-integer OAM value $(m-n)\sin2\alpha$. However, for the diagonal matrix elements ($a_{11}$ and $a_{22}$), corresponding to the angular shifts ($\Theta_\mathrm{GH}$ and $\Theta_\mathrm{IF}$), the results for the non-integer case are essentially different from the integer case. Apart from the contribution of $|m-n|$ to the shift, that is similar to the integer case \cite{Merano:PRA10}, there is an extra factor $\left(2\pm\cos2\alpha\right)/2$ showing the beam shift dependency on the angle $\alpha$. This leads to a \textit{nonlinear} interpolation between the integer OAM shifts.

 \section{Experiment}
 \label{sec:set-up}
\begin{figure}[htbp]
\begin{center}
\includegraphics{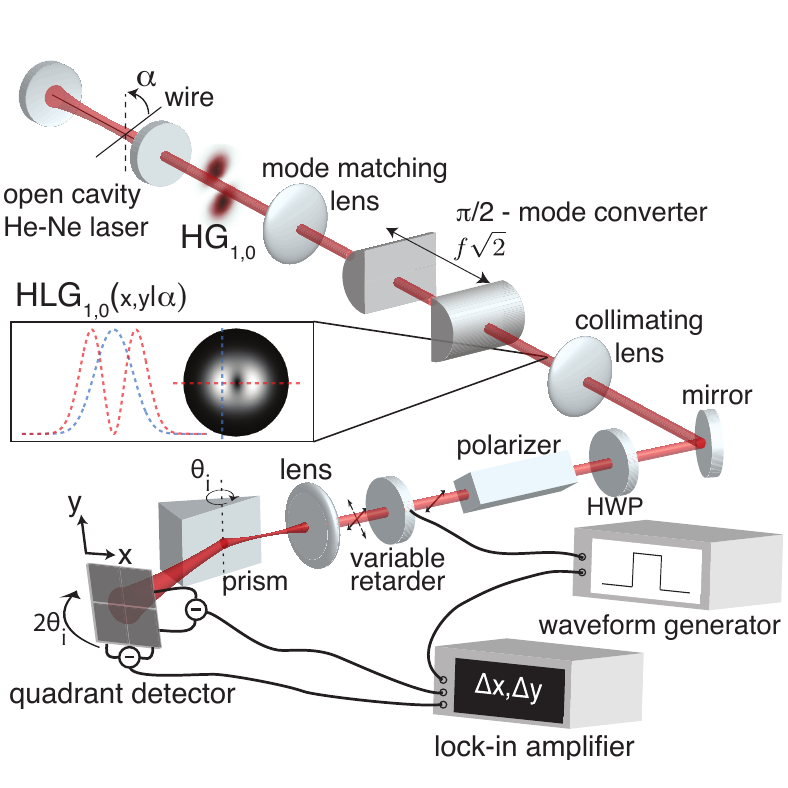}
\end{center}
\caption{\label{Fig02}Set-up for measuring non-integer OAM beams. Inset: a typical HLG beam with the intensity profile cross sections along the detector transverse axes.}
\end{figure}
To demonstrate the nonlinear interpolation of the angular beam shifts for non-integer OAM beams at partial external reflection, we use the set-up shown in Fig.~\ref{Fig02}. A HLG$_{1,0}$ beam is generated by modifying the orientation angle $\alpha$ of the incoming HG$_{1,0}$ mode with respect to the `$\pi/2$-mode converter'. The converter consists of a mode matching lens and a pair of cylindrical lenses separated at the appropriate distance \cite{Woerdman:OptCommun1993}. The generated beam is collimated and by using a half waveplate and a polarizer we tune the intensity and the input polarization, respectively. This beam is reflected from the hypothenusa air-glass interface of a glass prism ($n_2=1.51$). To measure the angular beam shift, we use a quadrant detector and obtain the relative beam displacement between two orthogonal directions of the incoming polarization states. Our generated HLG beams have the same symmetry beam axes (see inset of Fig.~\ref{Fig02}) as those of a standard quadrant detector, which greatly simplifies the beam shift measurement \cite{Nugrowati:GenDetFOAM2012}. To maximize the observable shift, we switch the polarization state of the beam between $0^o/90^o$ for $\Theta_\mathrm{GH}$ and $-45^o/45^o$ for $\Theta_\mathrm{IF}$, by means of a variable retarder. We further focus the beam to reach the condition for purely angular beam shifts to occur, i.e. $z\gg z_\mathrm{R}$ \cite{Aiello:NJP2012}. The angle of incidence at the air-glass interface is fixed at $\theta=45^o$ and we read the displacement signal of the reflected beam by using a lock-in amplifier. 

For completeness, we would like to point out that a quadrant detector furnishes the \textit{median} instead of the \textit{centroid} of  the beam intensity distribution \cite{Merano:PRA10, Aiello:NJP2012}. This leads to an additional factor in the matrix elements $a_{ij}$ for beam profiles that does not have the fundamental Gaussian distribution. We avoid this 'problem' when using a quadrant detector, by choosing HLG beams that have a Gaussian cross section (see the inset of Fig.~\ref{Fig02}) along the measured shifts \cite{Nugrowati:GenDetFOAM2012}. 

\begin{figure}[htbp]
\begin{center}
\includegraphics{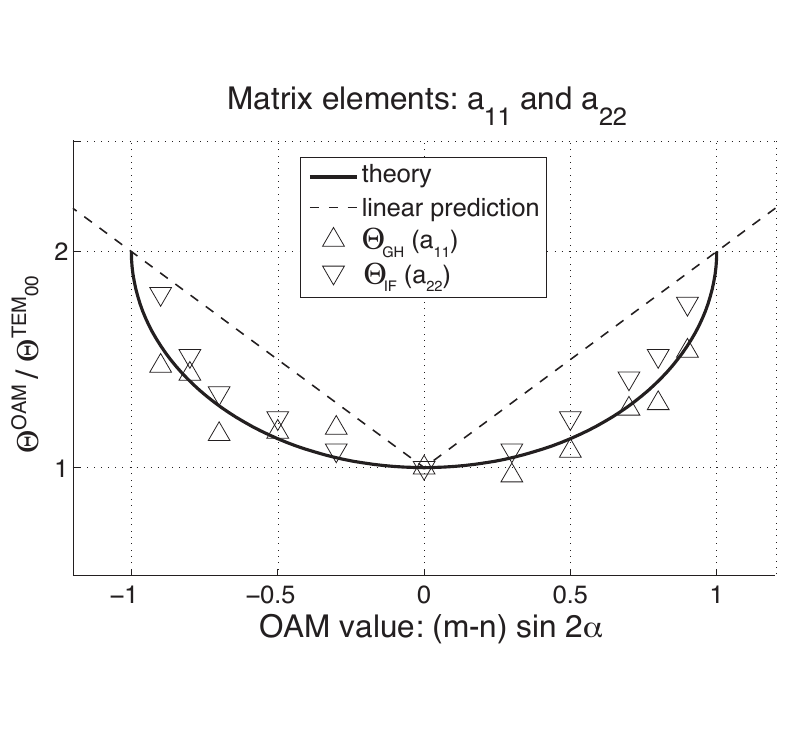}
\end{center}
\caption{\label{Fig03}Result of the angular shifts for HLG$_{1,0|\alpha}$ carrying non-integer OAM value -1\textless$\ell$\textless, normalized to that for TEM$_{00}$.}
\end{figure}

\begin{figure}[htbp]
\begin{center}
\includegraphics{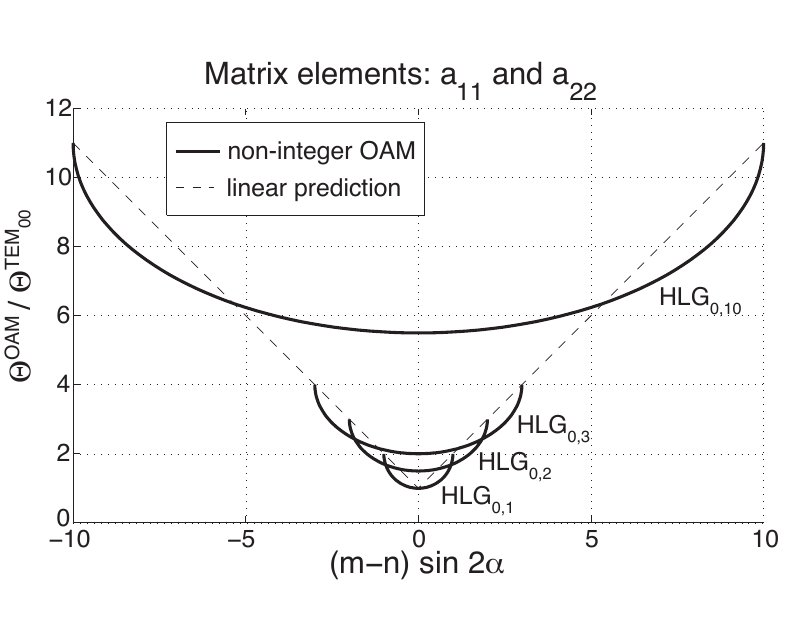}
\end{center}
\caption{\label{Fig04}Calculated angular shifts of higher order non-integer OAM beams.}
\end{figure}

Our result, shown in Fig.~\ref{Fig03}, clearly demonstrates the \textit{nonlinear} dependence of angular beam shifts on the non-integer OAM value. The data follows the theoretical prediction given in Eqs.(\ref{m11}) and (\ref{m22}). Due to the chosen HLG beams orientations, as discussed in the previous paragraph, we obtain the same curve for both $\Theta_\mathrm{GH}$ and $\Theta_\mathrm{IF}$. It is straightforward to generalize our two-component decomposition method in Eq.(\ref{eq:HLG10}) to any arbitrary non-integer OAM values carried by any higher order HLG beams. This is illustrated in Fig.~\ref{Fig04}, which shows the nonlinear interpolation of the beam shifts between higher-order integer OAM values. 

 \section{Conclusion}
 \label{sec:conclusion}
In conclusion, we have reported theoretically and experimentally, the angular GH and IF shifts of HLG beams carrying \textit{non-integer} OAM values. HLG beams are chosen to fulfill the requirement for meaningful beam shifts, in which the beam should be structurally stable upon propagation, apart from the overall scaling due to diffraction. 

Previous work has reported that the beam shifts in the external reflection case are linearly dependent on the integer OAM value \cite{Hermosa:SPIE2011}; it would seem natural that for non-integer OAM case, one should interpolate linearly between the integer OAM beam shifts. We have found, however, that the interpolation of beam shifts to non-integer OAM values by using HLG beams is highly nonlinear. Not only the OAM value, but also the OAM density \cite{Caravaca:PIERS2011} affect the overall beam shifts; the latter is more pronounced for non-integer OAM beams and the nonlinearity depends on the chosen mode profile. 

\section*{Acknowledgments}
This work is supported by the EU within FET Open-FP7 ICT as part of STREP Program 255914 Phorbitech. We acknowledge the fruitful discussions with Andrea Aiello, Sumant Oemrawsingh, and Steven Habraken.



\bibliographystyle{elsarticle-num}
\biboptions{sort&compress}
\bibliography{BeamShiftFOAM_2013}






\end{document}